\documentclass[
superscriptaddress,
final,
reprint,
showkeys,
pre,
aps,
floatfix,
]{revtex4-2}

\usepackage{tcolorbox}
\tcbset{
  width=1\textwidth,
  standard jigsaw,
  opacityback=0,   
}
\usepackage{pgffor}
\usepackage{graphicx}
\usepackage[utf8]{inputenc}
\usepackage{amsmath,amssymb,booktabs,latexsym,MnSymbol,bm,makecell}
\usepackage[%
  colorlinks=true,
  urlcolor=blue,
  linkcolor=blue,
  citecolor=blue
]{hyperref}
\usepackage[final]{pdfpages}
\makeatletter
\AtBeginDocument{\let\LS@rot\@undefined}
\makeatother
\begin{document}

\title{Causal asymmetry suggests productivity underlies scientific collaboration}

\author{Diego A. Frota}
\affiliation{Departamento de F\'isica, Universidade Federal do Cear\'a, 60451-970, Fortaleza, Cear\'a, Brazil}
\affiliation{Instituto Federal de Educa\c{c}\~ao, Ci\^encia e Tecnologia do Cear\'a, Sobral, Cear\'a, Brazil}

\author{Cesar I. N. Sampaio Filho}
\affiliation{Departamento de F\'isica, Universidade Federal do Cear\'a, 60451-970, Fortaleza, Cear\'a, Brazil}
\affiliation{Escola de Sa\'ude P\'ublica do Cear\'a, 60165-090, Fortaleza, Cear\'a, Brazil}

\author{V{\'i}tor H. Ribeiro}
\affiliation{Departamento de F\'isica, Universidade Estadual de Maring\'a, Maring\'a, PR, 87020-900, Brazil}

\author{Germano F. C. Luz}
\author{Humberto~A.~Carmona}
\affiliation{Departamento de F\'isica, Universidade Federal do Cear\'a, 60451-970, Fortaleza, Cear\'a, Brazil}
\affiliation{Escola de Sa\'ude P\'ublica do Cear\'a, 60165-090, Fortaleza, Cear\'a, Brazil}

\author{Matja{\v z}~Perc}\email{matjaz.perc@gmail.com}
\affiliation{Faculty of Natural Sciences and Mathematics, University of Maribor, Koro{\v s}ka cesta 160, 2000, Maribor, Slovenia}
\affiliation{Community Healthcare Center Dr. Adolf Drolc Maribor, Ulica talcev 9, 2000, Maribor, Slovenia}
\affiliation{Department of Physics, Kyung Hee University, Dongdaemun-gu, Seoul, 02447, Republic of Korea}
\affiliation{University College, Korea University, Seongbuk-gu, Seoul, 02841, Republic of Korea}

\author{Haroldo V. Ribeiro}\email{hvribeiro@uem.br}
\affiliation{Departamento de F\'isica, Universidade Estadual de Maring\'a, Maring\'a, PR, 87020-900, Brazil}
\affiliation{Instituto de Ciências Matemáticas e de Computação, Universidade de São Paulo, São Carlos, SP, Brazil}

\author{Jos\'e S. Andrade Jr.}\email{soares@fisica.ufc.br}
\affiliation{Departamento de F\'isica, Universidade Federal do Cear\'a, 60451-970, Fortaleza, Cear\'a, Brazil}
\affiliation{Escola de Sa\'ude P\'ublica do Cear\'a, 60165-090, Fortaleza, Cear\'a, Brazil}

\date{\today}

\begin{abstract}
Scientific productivity and collaboration are closely related, yet the causal direction between these two dimensions of academic careers remains unclear. Here we analyze the careers of 26,876 Brazilian researchers holding a nationally prestigious fellowship across multiple disciplines to test whether productivity shapes collaboration or the reverse. Focusing on total publications and distinct coauthors, we assess their directional relationship using a causal discovery framework based on asymmetries in conditional and distributional structure to determine the more plausible generative direction. Collaboration conditioned on productivity exhibits a highly regular mean-variance relation consistent with a geometric model, whereas the reverse conditioning yields a much less coherent pattern. Consistently, publication counts follow a lognormal distribution, whereas the coauthor distribution is reproduced by a geometric-lognormal mixture where productivity determines the expected number of collaborators. Surrogate tests indicate that the data are considerably more compatible with productivity shaping collaboration than with the reverse. The two quantities also accumulate asymmetrically: publication inequality rises steadily with career age, whereas coauthor inequality does not, indicating that advantage compounds more in output than in network breadth. Area-resolved analyses reveal a more nuanced picture, and although the direction where collaboration shapes productivity is never favored, the two directions are most clearly distinguished in areas more organized around laboratory and research-group models, where productivity inequality also grows most clearly. Among this cohort of prominent scientists, larger collaboration networks are thus more likely to reflect cumulative advantage in productivity than to independently generate greater output.
\end{abstract}

\maketitle

\begin{tcolorbox}
{\onecolumngrid
\noindent\textbf{Significance statement}
Scientific collaboration is often promoted as a direct route to greater research productivity, but the causal direction behind this association remains uncertain. Analyzing the careers of 26,876 leading Brazilian researchers, we find the data to be more consistent with productivity shaping collaboration than with the reverse. Publication counts show a regular structure from which the broad distribution of coauthorship can be reproduced, and inequality accumulates over careers in output rather than in network size. Large collaboration networks may therefore reflect cumulative advantage, as productive scientists become more visible and attract more collaborators. The evidence is sharpest in laboratory-based fields and inconclusive elsewhere, though it never favors the reverse. Encouraging collaboration alone may thus not necessarily increase scientific output.
}
\end{tcolorbox}
\vspace{3em}
\clearpage
\twocolumngrid

\section*{Introduction}
Scientific productivity and collaboration are two fundamental dimensions of academic careers, and their relationship has long attracted attention. Nearly a century ago, Lotka showed that productivity is highly unequal across researchers through his inverse-square law~\cite{lotka1926frequency}. Turning to collaboration, Price and Beaver~\cite{de1966collaboration} argued that collaborative networks, or their ``invisible colleges,'' help sustain this unequal output, with a small core of researchers driving most progress while coordinating a large floating population of less frequent coauthors. The pattern they observed---that prolific scientists also tend to be highly collaborative---was later documented in several contexts, including Zuckerman's study of Nobel laureates~\cite{zuckerman1967nobel}, Beaver and Rosen's historical analyses of the rise of scientific collaboration~\cite{de1979studies}, and Pravdi{\'c} and Olui{\'c}-Vukovi{\'c}'s study of repeated collaborative ties~\cite{pravdic1986dual}.

These early insights have since been explored at much larger scales through extensive analyses of collaboration networks, revealing how scientists connect through coauthorship and how these ties affect the production and diffusion of knowledge~\cite{newman2001structure, newman2001scientific, barabasi2002evolution, guimera2005team, jones2008multi, wang2013quantifying, fortunato2018science, wang2021science}, as well as the innovative character, impact, and interdisciplinarity of the resulting contributions~\cite{uzzi2013atypical, wu2019large, zeng2021fresh, liu2024scholars, tang2024investigating, ren2025collaborating, yang2025repeat}. Overall, these studies provide evidence of a positive association between productivity and collaboration. This association is also plausible on mechanistic grounds: collaboration enables scientists to combine complementary expertise and share labor, equipment, funding, and data~\cite{adams2005scientific, freeman2014why}, thereby potentially increasing research output, whereas highly productive scientists tend to accumulate visibility and reputation~\cite{merton1968matthew, agrawal2014collaboration}, making them more attractive collaborators and expanding their coauthorship networks. What remains much less clear is the causal direction underlying this relationship. Do researchers publish more because they collaborate more, or do they collaborate more because they publish more? Both mechanisms are plausible, and both may operate simultaneously. Yet the distinction matters. If collaboration causally increases productivity, then policies that stimulate team formation, mobility, or the expansion of coauthorship networks may enhance scientific output. If the dominant mechanism runs in the opposite direction, however, the benefits of collaboration may largely reflect cumulative advantage, whereby already productive scientists attract new collaborators. Clarifying this direction also has implications for how research systems should organize collaboration and allocate resources to support the accumulation and diffusion of useful knowledge, which have long been regarded as important foundations of economic and technological progress~\cite{mokyr2002gifts}.

Only a small number of studies have directly addressed this question using explicitly causal or endogeneity-aware approaches. Among these, Lee and Bozeman showed that the effect of collaboration on productivity can depend on how output is measured~\cite{lee2005impact}, whereas He and colleagues found that some forms of collaboration, particularly international collaborations, are associated with subsequent increases in research output~\cite{he2009research}. More directly, Ductor~\cite{ductor2015does} analyzed a panel of economists and found evidence that coauthorship exerts a positive causal effect on academic productivity. Conversely, Abramo \textit{et al.}~\cite{abramo2017relationship}, using a large longitudinal panel of Italian scholars, reported a more nuanced pattern in which prior productivity broadly stimulates subsequent collaboration, whereas only some forms of collaboration contribute positively to later productivity. Recent work by Zinilli and colleagues~\cite{zinilli2026collaborative} further examined the interplay between collaborative structures and research productivity at the university level using Bayesian networks, finding that higher productivity can increase universities' ability to attract partners in project-based networks in Physics and Engineering, while the relationship is less clear in Life Science. Together, these studies indicate that the productivity--collaboration association cannot be reduced to a simple correlation. Yet much of this literature focuses on broad forms of collaboration or institutional network position, rather than on the direct interplay between publication counts and total number of coauthors.

Here we address this gap by analyzing the scientific careers of more than 26,000 Brazilian researchers who held a nationally prestigious and highly competitive fellowship across multiple disciplines. We focus on two basic dimensions of scientific careers, namely, total publication output and total number of distinct coauthors, and investigate their directional relationship by adapting the causal framework introduced by Muchnik and colleagues~\cite{muchnik2013origins} within the broader perspective of causal inference~\cite{pearl2009causal}. This framework, which builds on nonlinear causal discovery methods~\cite{hoyer2008nonlinear, zhang2009identifiability, janzing2012information}, allows us to test whether heterogeneity in productivity generates the broad distribution of collaboration, or whether the reverse explanation is better supported by the data. Throughout this work, we use the term causal discovery in the specific technical sense established by this literature, in which the more plausible direction of dependence between two variables is inferred from asymmetries in their conditional and distributional structure. We emphasize that this approach is a deliberately lightweight form of causal inference, as it involves no temporal precedence, intervention, instrument, or counterfactual. Our results are thus not estimates of an intervention-level causal effect, but statements about which generative direction is more consistent with the population-level joint distribution of career totals.

Within this framework, we find a clear directional asymmetry: collaboration conditioned on productivity exhibits a highly regular mean-variance structure closely following the scaling expected from a simple geometric fluctuation model, whereas productivity conditioned on the number of collaborators shows a much weaker and less coherent pattern. Consistent with this asymmetry, publication counts follow a lognormal distribution, while the distribution of coauthors is accurately reproduced by a geometric-lognormal mixture, with productivity determining the expected number of collaborators. Surrogate tests reinforce this picture by showing that the empirical data are compatible with the hypothesis that productivity shapes collaboration but not with the reverse. A parallel asymmetry emerges in how the two quantities accumulate over careers. The inequality of publications within groups of researchers of similar career age rises steadily as careers advance, whereas the inequality of coauthors, although substantially higher in level, shows no comparable increase, indicating that cumulative advantage compounds more strongly on output than on the size of a researcher's network. Resolving both analyses by broad area of knowledge reveals considerable variation across fields. The two directions are most clearly distinguished in Exact and Earth Sciences, Engineering, and Biological Sciences, broadly the areas organized around laboratory and research-group models, and precisely those in which the inequality of productivity grows markedly more clearly than that of collaboration. In the remaining areas neither direction can be sharply distinguished and the inequality patterns weaken or reverse, but in no area is the reverse direction favored, so that where the evidence is informative it points to productivity and where it is not it is silent rather than contrary. Taken together, these findings indicate that, in this cohort of Brazilian fellowship holders, the broad heterogeneity in collaboration is more plausibly inherited from heterogeneity in productivity than the reverse.

\section*{Dataset description}

Our results are based on a dataset comprising the publication records of 26,876 Brazilian researchers who held the nationally prestigious \textit{Bolsa de Produtividade em Pesquisa} (Research Productivity Fellowship) between 2001 and 2023, each with at least 10 published articles and a publication span of 10 years or more between their first and last paper. Awarded since the 1970s by the National Council for Scientific and Technological Development (CNPq, Conselho Nacional de Desenvolvimento Cient\'ifico e Tecnol\'ogico), this competitive fellowship recognizes scholars distinguished for sustained and high-quality research output. The evaluation, however, is not based on publication counts alone. It weighs the impact of a researcher's output, measured through citation-based indicators, alongside the development of human resources through graduate and postdoctoral supervision, international insertion, project leadership, and contributions to the research system more broadly, with the criteria and their relative weights established by area-specific advisory committees. The fellowship therefore selects researchers not on productivity alone, but on a more complex combination of attributes, several of which may plausibly be reflected in a researcher's collaboration patterns as well as in their output. The two further restrictions we impose are mild and retain 96\% of all fellowship holders over the period analyzed, so that the fellowship is effectively the only selection applied. These researchers are drawn from diverse disciplines and institutions distributed across the entire national system, so that the cohort may be regarded as elite without being narrow in disciplinary or geographic terms. Our sample includes researchers from Exact and Earth Sciences (5,907), Biological Sciences (4,776), Engineering (3,331), Humanities and Human Sciences (3,301), Health Sciences (3,268), Agricultural Sciences (3,223), Applied Social Sciences (1,973), Linguistics, Literature and the Arts (999), and other areas (98), thus combining broad disciplinary coverage with greater concentration in science, technology, engineering and mathematics (STEM) and health-related fields. 

We obtained the publication records of these scholars---totaling 2,433,673 articles---from the \textit{Plataforma Lattes}, a curriculum platform that serves as the \textit{de facto} repository of academic profiles in Brazil and is regularly updated by active researchers, particularly fellowship holders, to document their publications and other research achievements. Increasingly used in science-of-science studies~\cite{araujo2014collaboration, araujo2017gender, sunahara2021association, sunahara2023universal, ribeiro2026gender}, this platform avoids name-disambiguation issues often present in other bibliographic data sources because the information is entered directly by researchers, making it particularly useful for investigating the association between productivity and collaboration. Additionally, focusing on this fellowship cohort allows us to examine patterns within a relatively homogeneous population, thereby providing a setting in which productivity and collaboration are particularly pronounced. This homogeneity, however, also delimits the scope and generality of our conclusions.

\section*{Methodology and results}

\begin{figure}[ht!]
   \centering
   \includegraphics[width=1\linewidth]{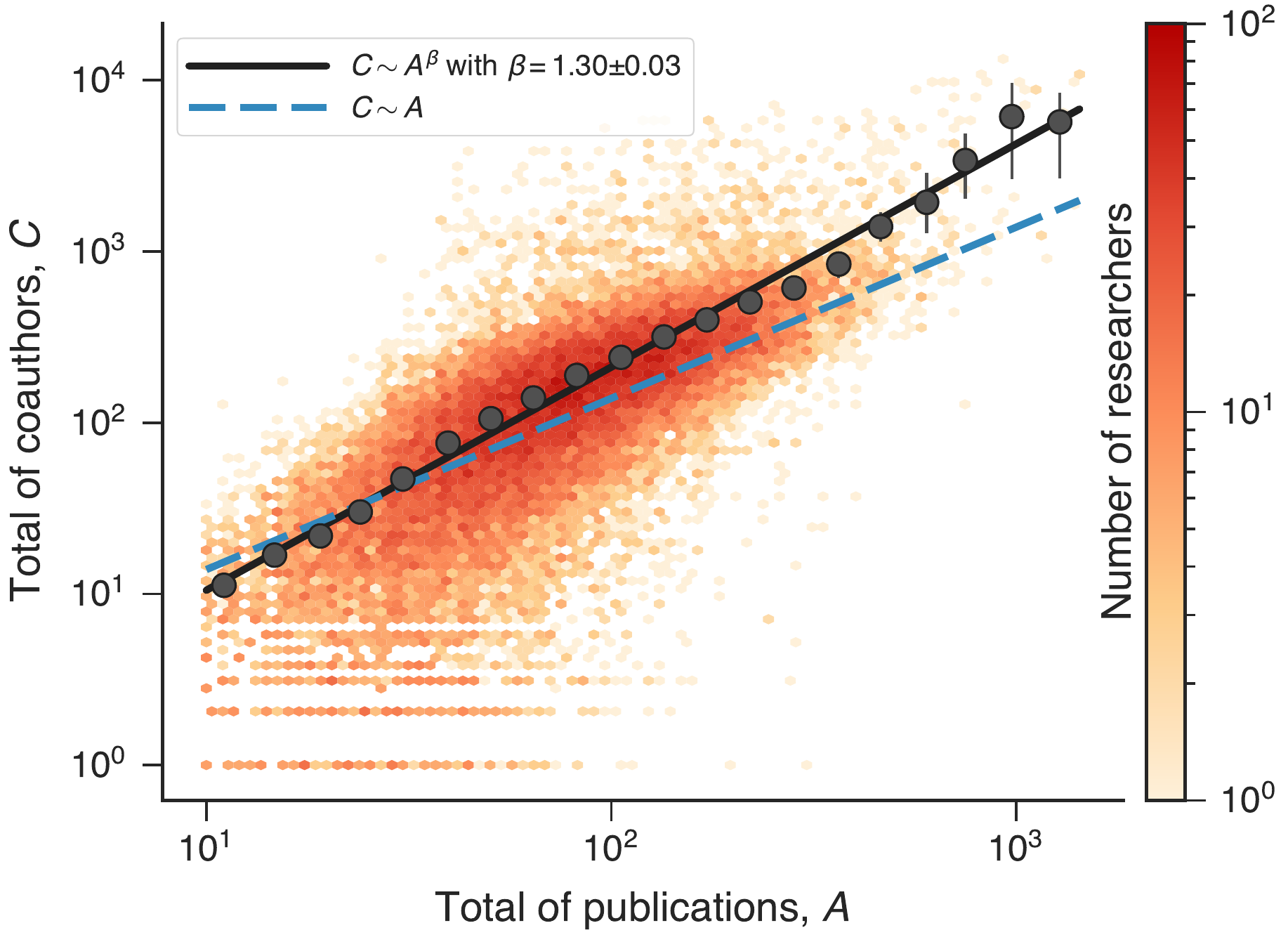}
   \caption{Relationship between scientific productivity and collaboration. Hexagonal binning map of the joint distribution of the total number of unique coauthors, $C$, and the total number of publications, $A$, for all researchers in our dataset. Both axes are shown on logarithmic scales, and hexagons are color-coded by the number of researchers in each bin. Gray circles denote the average value of $C$ within logarithmically spaced bins of $A$, with error bars representing 99\% bootstrap confidence intervals. The solid black line shows the best-fitting power-law relation, $C\sim A^\beta$, with $\beta=1.30\pm0.03$, whereas the dashed blue line indicates the linear scaling reference ($C\sim A$). We observe a strong positive correlation between collaboration and productivity, with the average number of coauthors increasing monotonically and superlinearly with publication output, indicating that gains in productivity are accompanied by a more than proportional increase (superlinear) in the number of collaborators.\vspace*{-0.6cm}}
   \label{fig:1}
\end{figure}

Using the obtained curricula, we summarize each researcher's activity and collaboration by counting the total number of publications ($A$) and the total number of unique coauthors ($C$). The number of unique coauthors is calculated by aggregating the author lists of all articles associated with each researcher after standardizing each coauthor's name into an identifier defined by the surname and the first initial of the given name, with diacritics, spaces, punctuation, common Portuguese particles (such as \textit{de} and \textit{dos}), and generational suffixes (such as \textit{Filho} and \textit{Junior}) removed. Although this procedure may occasionally merge distinct names sharing the same surname and initials or split variants of the same name with different levels of abbreviation, the standardization is applied separately within each individual curriculum, so such errors do not propagate across the full dataset and are expected to remain rare within a given researcher's coauthor list. We emphasize that this count includes every individual listed as an author regardless of role, as the Plataforma Lattes records author lists without metadata on contributions or roles. Figure~\ref{fig:1} shows the joint distribution of the total number of coauthors and publications for all researchers, revealing a clear and strong positive association between these two variables (Pearson $r=0.74$ and Spearman $\rho=0.77$ for the log-transformed values of $A$ and $C$, both with $p$-value~$<0.001$). Furthermore, when conditioning on publications and averaging the number of coauthors, we find a smooth and monotonically increasing relationship that is well approximated by a power law, $C\sim A^\beta$, with $\beta=1.3$, which appears as a linear trend on the logarithmic scales of Figure~\ref{fig:1}. This thus shows that collaboration scales superlinearly with productivity, so that the number of distinct coauthors grows more than proportionally with publication output; specifically, each 1\% increase in productivity is associated, on average, with a 1.3\% increase in the number of coauthors. Conversely, when conditioning on the number of coauthors and averaging the total number of publications, we find a much less clear and more complex relationship (Figure~S1), already indicating an asymmetric behavior in the joint distribution of publications and coauthors.

\begin{figure*}[ht!]
  \centering
  \includegraphics[width=0.9\linewidth]{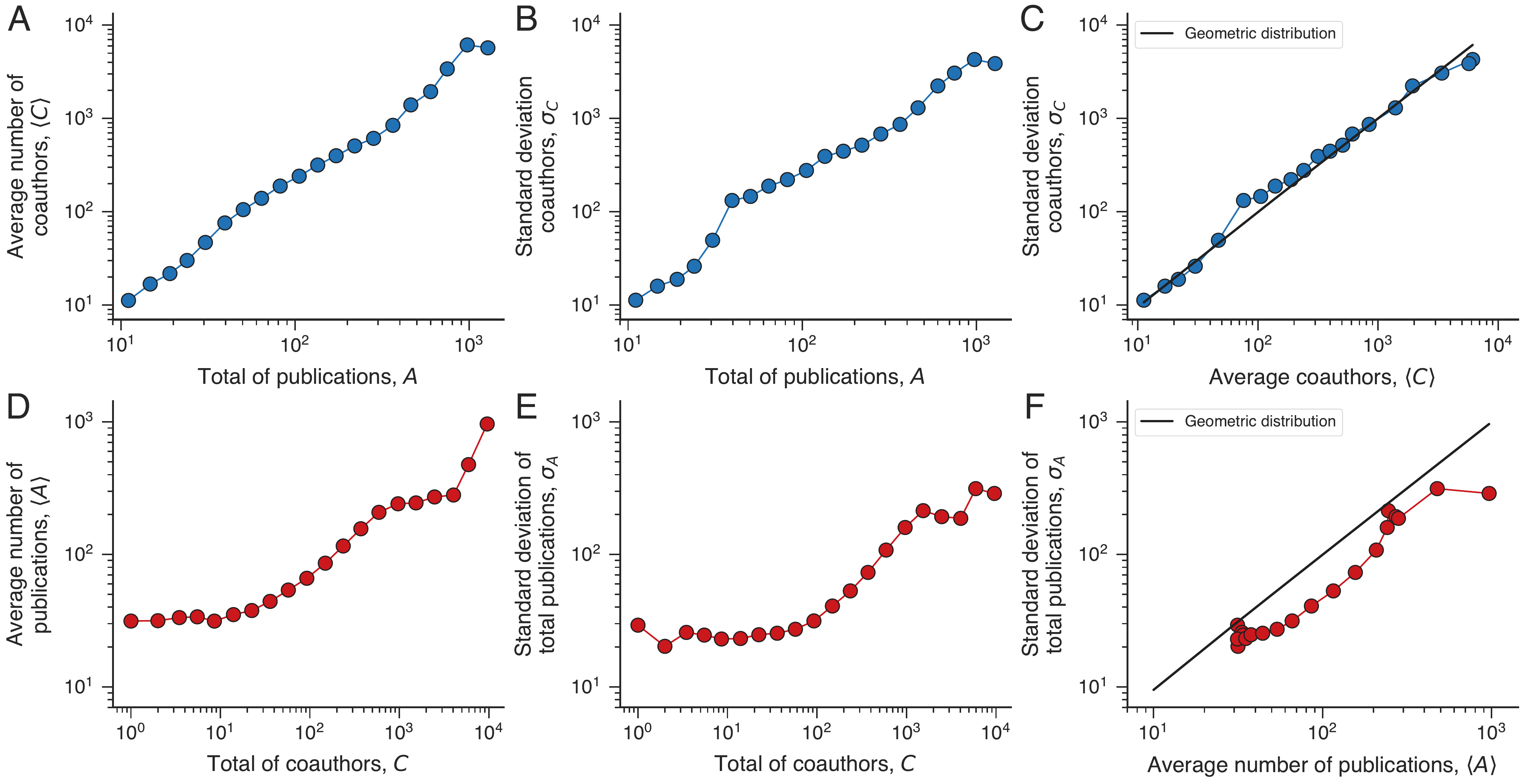}
  \caption{Conditional asymmetries in the joint distribution of productivity and collaboration. (A) Average number of coauthors $\langle C \rangle$ and (B) standard deviation of the number of coauthors $\sigma_C$, both as functions of the total number of publications $A$. (C) Relation between the standard deviation $\sigma_C$ and average number of coauthors $\langle C \rangle$, conditioned on the number of publications. Together, these statistics reveal a smooth and regular conditional structure for collaboration given productivity. (D) Average number of publications $\langle A \rangle$ and (E) standard deviation of the number of publications $\sigma_A$, both as functions of total number of coauthors $C$. (F) Relation between the standard deviation $\sigma_A$ and average number of publications $\langle A \rangle$ conditioned on the number of coauthors. In contrast to the conditional statistics obtained when conditioning on productivity, these results indicate that the statistics of productivity given collaboration are markedly less regular, highlighting an asymmetry in the joint distribution of $A$ and $C$. Markers in all panels represent averages computed within $20$ logarithmically spaced bins of the conditioning variable, and all axes are shown on logarithmic scales. The solid black lines correspond to the relation between mean and standard deviation expected for geometric distributions, namely, $\sigma_C = \sqrt{\langle C\rangle(\langle C\rangle-1)}$ and $\sigma_A = \sqrt{\langle A\rangle(\langle A\rangle-1)}$ in panels C and F, respectively, which are practically indistinguishable from a 1:1 relation over the observed range of mean values.\vspace*{-0.4cm}}
\label{fig:2}
\end{figure*}

Asymmetries in conditional distributions lie at the core of nonlinear causal discovery frameworks for evaluating competing directional hypotheses about the association between variables~\cite{hoyer2008nonlinear, zhang2009identifiability, janzing2012information}. Nonlinear dependencies typically break the symmetry between forward and reverse conditional distributions, and identifying which variable exhibits more regular behavior can help infer which is more likely to act as the driving factor, especially because such asymmetries arise naturally when the functional relationship between variables is non-invertible~\cite{friedman2000gaussian, shimizu2006linear}. Building on these ideas, and following the approach of Muchnik \textit{et al.}~\cite{muchnik2013origins}, we investigate the conditional structure of the joint distribution of publications and coauthors by calculating conditional averages and standard deviations under the two competing directional hypotheses that productivity drives collaboration ($A\to C$) and that collaboration drives productivity ($C\to A$). Figures~\ref{fig:2}A and \ref{fig:2}B depict the conditional average, $\langle C \rangle$, and standard deviation, $\sigma_C$, of the number of coauthors as functions of the total number of publications (calculated using 20 logarithmically spaced bins), indicating that both quantities increase smoothly and monotonically with publication output. Moreover, when conditioning on publications, Figure~\ref{fig:2}C shows that the relationship between the standard deviation and the average number of coauthors traces a smooth and narrow curve lying very close to the 1:1 line and thus in good agreement with the theoretical expectation for a geometric distribution. By contrast, conditioning on the total number of coauthors yields a markedly different picture. The conditional average, $\langle A \rangle$, and standard deviation, $\sigma_A$, of the number of publications, shown as functions of the total number of coauthors in Figures~\ref{fig:2}D and \ref{fig:2}E, indicate a weaker association characterized by a relatively flat behavior at small numbers of coauthors, particularly for the standard deviation of publications. As further shown in Figure~\ref{fig:2}F, this translates into a much more irregular relationship between standard deviation and the average number of publications conditioned on the number of coauthors, with systematic deviations from the geometric scaling approximated by the 1:1 relation. We also note that this qualitative pattern is highly robust to the number of bins used to calculate the conditional averages and standard deviations, both when conditioning on publications (Figure~S2) and on coauthors (Figure~S3).

Combined, these results show that collaboration conditioned on productivity exhibits a well-behaved structure, motivating a simple generative mechanism in which the average number of coauthors is determined by a monotonic function of productivity, $\langle C\rangle\sim A^\beta$, as observed empirically, and, for a given level of productivity, the number of coauthors follows a geometric distribution, namely,
\begin{equation}\label{eq:geometric}
p(C|\langle C\rangle) = (\langle C\rangle-1)^{C-1}\langle C\rangle^{-C}\,,
\end{equation}
thereby yielding the mean-standard deviation relation $\sigma_C = \sqrt{\langle C\rangle(\langle C\rangle-1)}$, which is closely followed by the data in Figure~\ref{fig:2}C, further motivating our geometric model. Conversely, the results for productivity conditioned on collaboration do not admit an equally simple description, as the corresponding conditional statistics are less regular, deviate from geometric scaling, and do not suggest a comparably natural generative form. This asymmetry is thus more consistent with a generative direction in which productivity shapes collaboration than with the reverse. It also motivates examining the marginal distributions of productivity and collaboration to assess whether their overall forms are more naturally explained by a mechanism in which heterogeneity in productivity generates the observed variability in collaboration, or the reverse.

\begin{figure*}[ht!]
   \centering
   \includegraphics[width=0.80\linewidth]{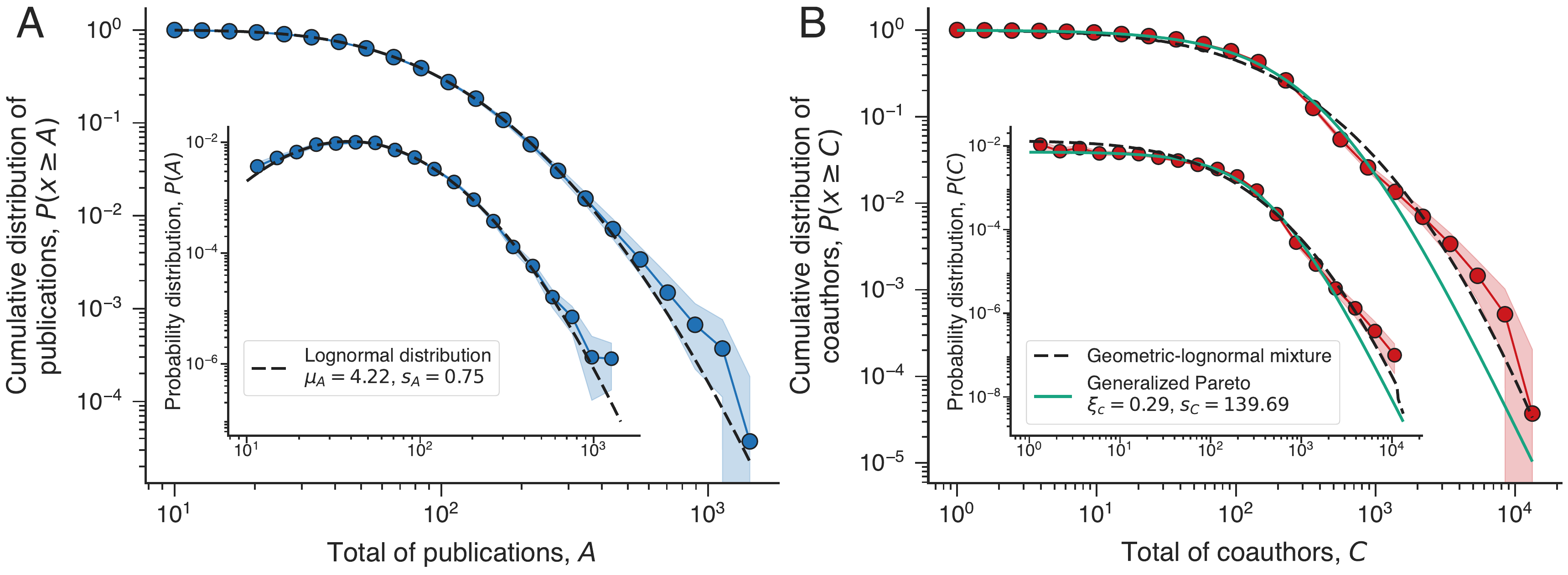}
   \caption{Marginal distributions of productivity and collaboration. (A) Complementary cumulative distribution of the total number of publications $A$ across all researchers. Shown together is the best-fitting lognormal distribution, Equation~\ref{eq:lognormal} with $\mu_A=4.22$ and $s_A=0.75$, represented by the dashed black line. The inset shows the corresponding probability distribution compared with the same lognormal fit. (B) Complementary cumulative distribution of the total number of coauthors $C$, shown together with the prediction of the geometric-lognormal mixture model, Equation~\ref{eq:lognormalmix}, represented by the dashed black line and, for comparison, the best-fitting generalized Pareto distribution, Equation~\ref{eq:genpareto} with $\xi_C=0.29$ and $s_C=139.69$, shown by the solid green line. The inset shows the corresponding probability distribution with the same fitted curves. In both panels, markers denote the empirical distributions, shaded bands indicate bootstrapped 95\% confidence intervals, and all axes are shown on logarithmic scales.\vspace*{-0.4cm}}
   \label{fig:3}
\end{figure*}

We thus estimate the marginal distribution of the total number of publications across all researchers, as shown in Figure~\ref{fig:3}A. Reflecting the strong heterogeneity in productivity already highlighted by Lotka's classical inverse-square law of scientific output~\cite{lotka1926frequency}, this distribution is highly right-skewed, spanning more than two orders of magnitude in $A$ and nearly five orders of magnitude in cumulative probability. In our data, however, we find that the empirical distribution of productivity is much better described by a lognormal than by a power-law distribution. The dashed line in Figure~\ref{fig:3}A shows the best-fitting lognormal distribution, 
\begin{equation}\label{eq:lognormal}
   p(A) = \frac{1}{s_A A \sqrt{2\pi} }\exp\left[-\frac{(\ln A-\mu_A)^2}{2s_A^2}\right]\,,
\end{equation}
with parameters $\mu_A=4.22$ and $s_A=0.75$ estimated by maximum likelihood. We assess the quality of this fit using a parametric bootstrap Kolmogorov--Smirnov (KS) goodness-of-fit test by generating 10,000 synthetic samples from the fitted lognormal model, refitting the model at each iteration, and calculating the KS statistic for each sample. This analysis yielded a $p$-value of $0.46$, indicating that roughly half of the synthetic samples produced KS statistics larger than that observed in the empirical data, and thus that the productivity distribution is well consistent with the lognormal hypothesis. We also compare the lognormal with a broad set of alternative distributions (see Table~S1), and across all tests it consistently emerges as the best-fitting model. Beyond its strong statistical support, the lognormal description is also consistent with previous findings. In particular, earlier studies~\cite{shockley1957statistics, fronczak2007analysis, ruocco2017bibliometric} reported lognormal productivity patterns for relatively homogeneous groups of highly productive researchers, similar to the cohort analyzed here, although deviations in the upper tail have also been observed~\cite{cortes2016productivity}. At the same time, lognormality is consistent with mechanistic interpretations in which productivity arises from multiplicative processes operating across scientific careers, including cumulative advantage, preferential visibility, and resource-allocation dynamics~\cite{limpert2001log, sheridan2018preferential, petersen2014inequality}.

In the case of the marginal distribution for the number of coauthors, Figure~\ref{fig:3}B shows that it is also highly right-skewed, spanning over four orders of magnitude in $C$ and nearly five orders of magnitude in cumulative probability, with a tail that decays more slowly than that of the productivity distribution. Given that the conditional statistics of collaboration support the idea that the average number of coauthors tracks researchers' productivity and that, conditioned on the productivity, the distribution of coauthors is compatible with a geometric form, we model the marginal distribution of coauthors as a geometric-lognormal mixture given by
\begin{equation}\label{eq:lognormalmix}
    p(C) = \int p(C | \langle C \rangle) p(A) dA\,,
\end{equation}
where $\langle C \rangle= b A^\beta$ is a power-law function of productivity $A$, with $\beta=1.30$ and $b=0.33$, $p(C | \langle C \rangle)$ is the geometric distribution of Equation~\ref{eq:geometric}, and $p(A)$ is the marginal distribution of productivity given by Equation~\ref{eq:lognormal}, with $\mu_A=4.22$ and $s_A=0.75$. Although the integral in Equation~\ref{eq:lognormalmix} has no simple closed form, it can be easily evaluated numerically, yielding a functional form that can be directly compared with the empirical distribution without introducing additional fitting parameters, since both the parameters of the power-law relation between $\langle C \rangle$ and $A$ as well as those of the lognormal distribution of $A$ are independently determined from the data. 

The dashed line in Figure~\ref{fig:3}B indicates that the resulting geometric-lognormal mixture is in good overall agreement with the empirical distribution, despite a systematic deviation observed in the tail, where the curve falls outside the confidence intervals of the empirical estimates and tends to underestimate the number of coauthors. We remark, however, that no fitting was performed on the coauthor data, as the geometric-lognormal mixture follows entirely from the conditional analysis together with the independently estimated productivity distribution. Moreover, much of the discrepancy can be attributed to a subtle inflection point slightly below $C \sim 10^3$, a feature that is difficult to capture with standard parametric forms. Indeed, we find that a broad range of alternative distributions is incompatible with the coauthor data (Table~S2), with the generalized Pareto distribution 
\begin{equation}\label{eq:genpareto}
   p(C) = \frac{1}{s_C}\left(1+\xi_C \frac{C}{s_C}\right)^{-1-\frac{1}{\xi_C}}\,,
\end{equation}
where $\xi_C$ and $s_C$ are fitting parameters, emerging as the best-performing parametric form. However, as also shown in Figure~\ref{fig:3}B by the solid green line, the generalized Pareto exhibits deviations from the empirical behavior that are particularly pronounced in the upper tail. These results indicate that the geometric-lognormal mixture captures the main structure of the coauthor distribution remarkably well without direct fitting, which is consistent with the broad heterogeneity in collaboration arising largely from underlying differences in productivity combined with geometric fluctuations conditioned on it, and hence with the generative direction in which productivity shapes collaboration.

\begin{figure}[ht!]
\centering
\includegraphics[width=0.90\linewidth]{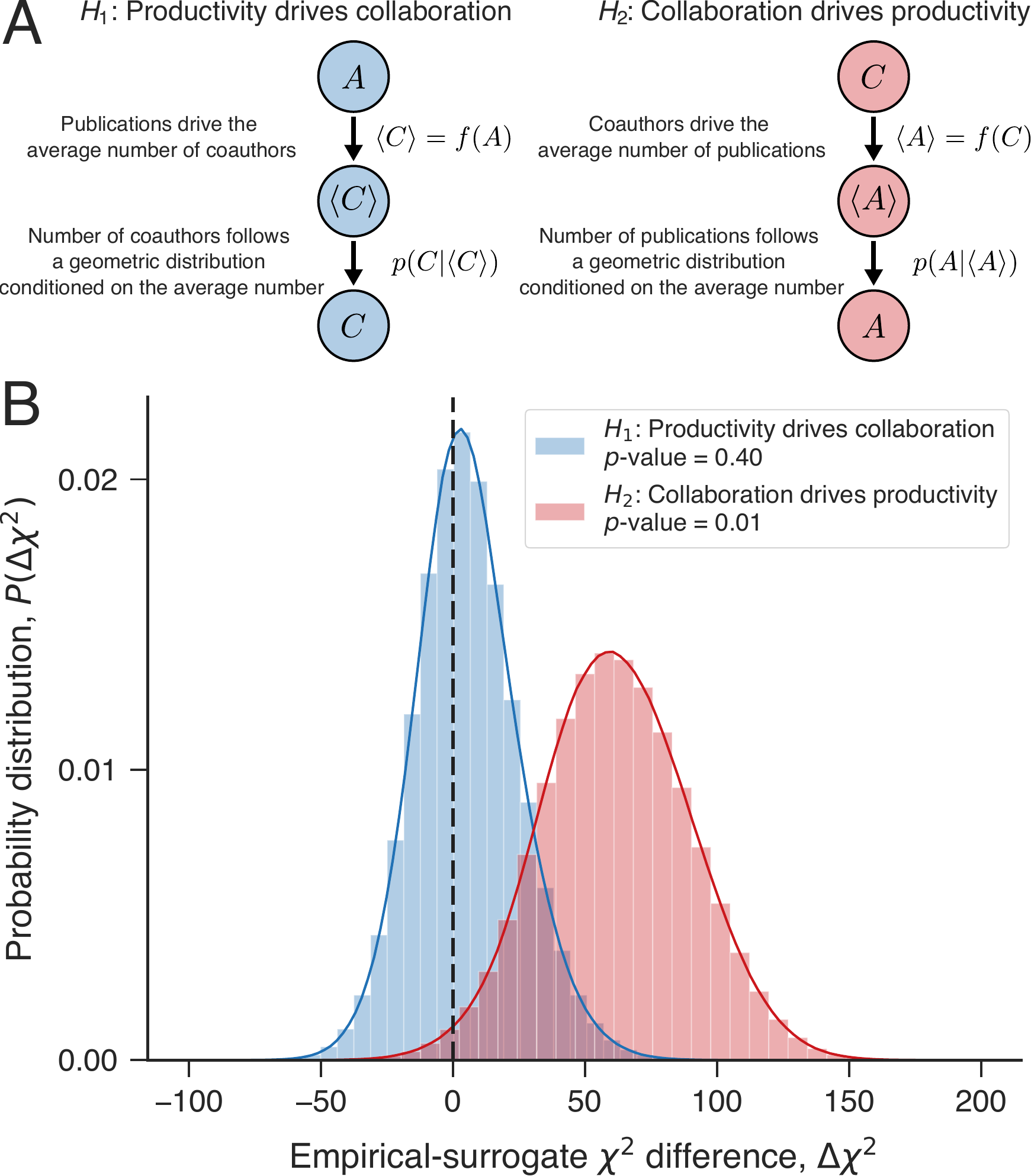}
\caption{Surrogate-based directional test of the productivity--collaboration relationship. (A) Schematic representation of the two competing generative hypotheses. Under $H_1$ (left), productivity drives collaboration, so that the total number of publications $A$ sets the average number of coauthors via some function $f$, $\langle C\rangle=f(A)$, from which the observed number of coauthors $C$ is drawn according to a geometric distribution. Under $H_2$ (right), collaboration drives productivity, so that the total number of coauthors $C$ sets the average number of publications via some function $f$, $\langle A\rangle=f(C)$, from which the observed number of publications $A$ is drawn according to a geometric distribution. (B) Distributions of the empirical-surrogate $\chi^2$ difference, $\Delta\chi^2=\chi^2_{\mathrm{emp}}-\chi^2_{\mathrm{surr}}$, obtained from the surrogate test for the two directional hypotheses using $100$ percentile-based bins. For $H_1$, the distribution is centered near zero and yields a $p$-value of $0.40$, whereas for $H_2$ it is shifted towards positive values and yields a $p$-value of $0.01$. These results favor $H_1$ over $H_2$ as the generative direction more consistent with the data.\vspace*{-0.6cm}
}
\label{fig:4}
\end{figure}

We further perform a surrogate-based directional analysis of the two competing generative hypotheses, $H_1$ and $H_2$, schematically illustrated in Figure~\ref{fig:4}A. The main purpose of this test is not to introduce a distributional hypothesis, but to quantify, beyond visual comparison, the asymmetries revealed by our preceding results. Under $H_1$, we assume that productivity drives collaboration, so that the total number of publications $A$ determines the average number of coauthors through a function, $\langle C \rangle = f(A)$, after which the observed number of coauthors $C$ is drawn from a geometric distribution conditioned on $\langle C \rangle$. Under $H_2$, we assume that collaboration drives productivity, so that the total number of coauthors $C$ determines the average number of publications through a function, $\langle A \rangle=f(C)$, after which the observed number of publications $A$ is drawn from a geometric distribution conditioned on $\langle A \rangle$. For each directional hypothesis, we partition the candidate cause variable into percentile-based bins and, within each bin, record the empirical values of the corresponding effect variable, compute their mean, and generate surrogate datasets by drawing geometric random variables with that same mean. We then calculate the $\chi^2$ statistic between the empirical conditional effect values and corresponding surrogate values ($\chi^2_{\mathrm{emp}}$), as well as the $\chi^2$ statistic between two independent surrogate realizations drawn from the same null model ($\chi^2_{\mathrm{surr}}$), and define their difference $\Delta\chi^2=\chi^2_{\mathrm{emp}}-\chi^2_{\mathrm{surr}}$. By considering the values of $\Delta\chi^2$ across all bins and repeating the entire procedure $10^5$ times, we estimate the distribution of empirical-surrogate $\chi^2$ difference under each generative hypothesis. 

If a directional hypothesis is a plausible generative model for the data, the empirical conditional structure should be reproduced by its surrogate model, yielding $\Delta\chi^2$ values centered near zero; conversely, systematic positive deviations indicate that the proposed causal direction does not reproduce the conditional structure observed in the data. As a summary of the test, we also calculate a one-sided $p$-value defined as the fraction of realizations for which  $\Delta\chi^2<0$, corresponding to the fraction of cases in which the empirical discrepancy is smaller than the discrepancy observed between two independent surrogate realizations. High $p$-values indicate compatibility between the empirical data and the surrogate model, whereas low $p$-values indicate that the proposed direction is inconsistent with the data. We note that this procedure involves no adjustable parameters, as the geometric distribution is fully determined by its mean, which is calculated directly from the data within each bin. Both directions are thus measured against the same minimal reference, and the asymmetry we report is precisely that this reference reproduces the conditional structure of $C$ given $A$ but fails to reproduce that of $A$ given $C$. A more flexible family, such as one carrying a free dispersion parameter, would instead allow the surrogate models to absorb that asymmetry into their fitted parameters rather than quantify it.

The results of the surrogate test with 100 percentile-based bins shown in Figure~\ref{fig:4}B further favor the direction in which productivity shapes collaboration. Indeed, we find that under $H_1$, the distribution of $\Delta\chi^2$ is centered near zero, indicating that the empirical conditional structure of collaboration given productivity is well reproduced by the surrogate model. Consistently, the associated one-sided $p$-value is $0.40$, showing that the empirical discrepancy is frequently comparable to, or smaller than, that expected between independent surrogate realizations. By contrast, under $H_2$, the corresponding distribution of $\Delta\chi^2$ is shifted towards positive values and centered around $\Delta\chi^2\approx60$, indicating that the empirical conditional structure of productivity given collaboration is much less compatible with the surrogate model implied by this direction. Accordingly, the associated $p$-value is only $0.01$, meaning that cases in which the empirical discrepancy is smaller than the surrogate discrepancy are rare. Importantly, this conclusion is not sensitive to the specific binning choice as the same scenario is obtained across a broad range of bin numbers (Figure~S4). 

The observed asymmetries indicate which generative direction the data favor but are uninformative about the mechanisms that likely produced them. One natural candidate is cumulative advantage~\cite{merton1968matthew, price1976general}, the general principle that success breeds further success in scientific careers. Two features of our results are consistent with a mechanism of this kind: the lognormal form of the productivity distribution carries the signature of a multiplicative process, and collaboration scales superlinearly with output, so that each additional publication brings a more than proportional increase in the expected number of collaborators. Cumulative advantage can, however, operate on either quantity. If advantage compounds primarily on productivity, sustained output would earn recognition, resources, and scientific visibility, making productive researchers more attractive as collaborators and expanding the pool of those who seek to coauthor with them, so that much of the breadth of a collaboration network would follow from scientific output. If, instead, cumulative advantage operates mainly through collaboration, each new tie would provide access to additional collaborators, resources, and opportunities to publish, so that much of a researcher's output would follow from expanding their networks. These possibilities are not mutually exclusive, as advantage may compound in both quantities, and each predicts skewed distributions and a strong association between productivity and collaboration, so invoking cumulative advantage does not by itself settle the direction. However, if this cumulative advantage acts more strongly on one quantity than on the other, it can leave a detectable trace over the course of a career.

Allison and Stewart~\cite{allison1974productivity} proposed a criterion for detecting such a trace in cross-sectional data such as ours. Their argument is that cumulative advantage implies that the gains (additional publications or coauthors) increase with the level already attained. Researchers at similar career stages thus diverge progressively as their careers advance, leaving an observable signature in the form of increasing within-cohort inequality. Following their approach, we group researchers by career age (measured as years since PhD) and treat the resulting groups as approximations of cohorts observed at different career stages, allowing us to compare their levels of inequality. Although these groups do not represent the same cohort followed longitudinally, the variable on which cumulative advantage operates more strongly should exhibit a clearer increase in inequality across groups of increasing career age. We test this expectation by calculating the Gini index~\cite{allison1978measures} of the numbers of publications and coauthors within each group (Figure~S5). The publication Gini rises monotonically and approximately linearly with career age ($R^2=0.97$, at a rate of $0.019$ per decade), qualitatively reproducing the signature documented by Allison and Stewart. Collaboration inequality is substantially higher in level (Gini approximately $0.6$ against $0.4$), so collaboration is the more unequally distributed of the two quantities, yet it increases far less regularly ($R^2=0.50$) and non-monotonically, decreasing between $10$ and $20$ years, increasing between $20$ and $30$ years, and then saturating. The signature of cumulative advantage is therefore considerably stronger for productivity than for collaboration, a pattern that provides additional evidence on the direction of the productivity--collaboration relationship, independent of the conditional structure examined previously.

\begin{figure}[ht!]
\centering
\includegraphics[width=1\linewidth]{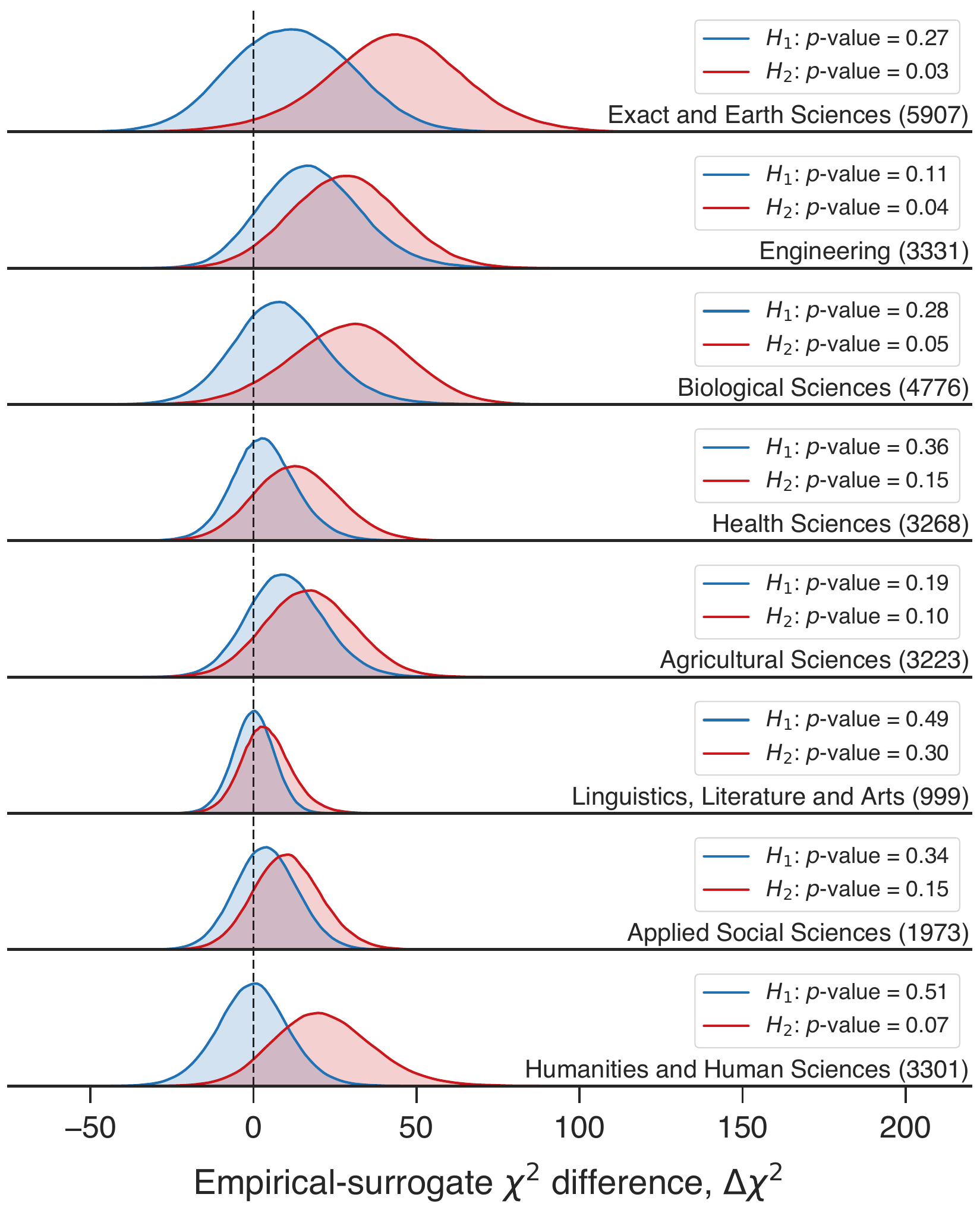}
\caption{Surrogate-based directional test resolved by broad area of knowledge. Distributions of the empirical-surrogate $\chi^2$ difference, $\Delta\chi^2=\chi^2_{\mathrm{emp}}-\chi^2_{\mathrm{surr}}$, obtained from the surrogate test for the two competing directional hypotheses within each broad area of knowledge, with the area and the corresponding number of researchers indicated in each panel. The blue distributions correspond to $H_1$ (productivity drives collaboration, $A\rightarrow C$) and the red distributions to $H_2$ (collaboration drives productivity, $C\rightarrow A$), with the corresponding $p$-values annotated in each panel. In every area, the distribution under $H_1$ remains centered near zero, whereas that under $H_2$ is systematically shifted towards positive values. Accordingly, $H_1$ is never rejected, while $H_2$ is rejected or borderline at the nominal 5\% level ($p$-values not adjusted for multiple comparisons) in Exact and Earth Sciences, Engineering, and Biological Sciences, and is not rejected in the remaining areas.\vspace*{-0.6cm}}
\label{fig:5}
\end{figure}

Areas of knowledge differ markedly in authorship conventions, publication rhythms, and collaboration cultures, which raises the question of whether the asymmetry observed in the aggregate manifests differently across them. We therefore repeat the surrogate tests separately within each broad area of knowledge, as shown in Figure~\ref{fig:5} (the 98 researchers assigned to other areas are not considered here). Across all areas, the tests never favor $H_2$: the distribution of $\Delta\chi^2$ under $H_1$ remains centered near zero and $H_1$ is never rejected, whereas the distribution under $H_2$ is systematically shifted towards positive values. There is, however, appreciable variation in the extent to which $H_2$ can be rejected. At the nominal 95\% confidence level, this hypothesis is rejected for Exact and Earth Sciences ($p$-value~$=0.03$) and Engineering ($p$-value~$=0.04$) and is borderline for Biological Sciences ($p$-value~$=0.05$), whereas it is not rejected for Health Sciences ($p$-value~$=0.15$), Agricultural Sciences ($p$-value~$=0.10$), Linguistics, Literature and the Arts ($p$-value~$=0.30$), Applied Social Sciences ($p$-value~$=0.15$), and Humanities and Human Sciences ($p$-value~$=0.07$), so that the analysis becomes inconclusive for these latter areas in the sense that neither direction can be sharply distinguished. This variation does not appear to reflect differences in the productivity of each area (Table~S3): for instance, Health Sciences exhibits the highest median number of publications (126 articles) and neither hypothesis is rejected, Exact and Earth Sciences presents an intermediate median (60 articles) and yields the clearest rejection of $H_2$, and Linguistics, Literature and the Arts combines the lowest median (36 articles) with the weakest evidence against $H_2$. Differences in collaboration culture provide a more plausible account of this variation, as the three areas in which $H_2$ is rejected or borderline are more organized around traditional laboratory or research-group models, where much of the coauthorship arises within an investigator's own group. The remaining areas appear to behave differently, and to better understand these patterns we repeat the inequality analysis separately within each area (Figure~S6).

These results indicate that the inequality of productivity grows with career age markedly more clearly than that of collaboration in areas where $H_2$ is rejected or borderline: for Exact and Earth Sciences, the publication Gini increases with years since PhD ($R^2=0.61$) while the coauthor Gini remains essentially flat ($R^2=0.04$), and the same pattern holds for Engineering ($R^2=0.91$ against $R^2=0.16$) and Biological Sciences ($R^2=0.97$ against $R^2=0.85$, both increasing but the former more strongly). This suggests that the cumulative advantage operates more strongly on productivity than on collaboration in these areas, consistent with collaboration being inherited from productivity rather than accumulating through an independent process. In most areas where $H_2$ cannot be rejected, the ordering weakens or reverses. In Applied Social Sciences, the coauthor Gini grows more clearly with career age ($R^2=0.66$) than the publication Gini ($R^2=0.17$), inverting the pattern observed in the traditionally laboratory-based fields. In Agricultural Sciences, collaboration inequality grows at least as clearly as productivity inequality ($R^2=0.85$ against $R^2=0.80$). In these two areas, collaboration thus appears to accumulate over careers partly independently of an individual's own output, which plausibly reflects more collective authorship practices and would account for the weaker asymmetry in our directional tests. In Linguistics, Literature and the Arts and in Humanities and Human Sciences, where collaboration is often more limited and irregular, neither inequality grows so clearly with career age ($R^2\leq0.31$ in all cases), suggesting that cumulative advantage in productivity and collaboration may be weaker or less consistently expressed in these areas, which is also consistent with the lack of clear asymmetry in our directional tests. Health Sciences is, however, an exception, as $H_2$ cannot be rejected but its inequality patterns are qualitatively similar to those of Exact and Earth Sciences and Engineering, with productivity inequality growing more strongly with career age ($R^2=0.96$) than collaboration inequality ($R^2=0.29$), indicating that the cumulative-advantage signal in productivity may persist even where the surrogate test is inconclusive. While we cannot fully account for this pattern, we note that health research is also largely organized around laboratories and large research groups, but this structure often coexists with multi-center and consortium authorship, a practice less common in the physical and engineering sciences. Such a more collective organization may add collaborators to an individual's records in a manner less tied to their own output, which in turn may blur the conditional asymmetry probed by our surrogate test while preserving the marginal accumulation of productivity inequality.

Overall, the surrogate-based directional test and the two inequality curves, resolved by broad areas of knowledge, reveal a more nuanced picture than the aggregate results alone suggest. The correspondence between the two analyses is closest in Exact and Earth Sciences, Engineering, and Biological Sciences, where $H_1$ is not rejected while $H_2$ is rejected or borderline, and the conditional asymmetries coincide with a marked dominance of productivity inequality. It is weaker in the remaining areas, where neither hypothesis can be rejected and the two inequality curves differ from area to area. This contrast should nonetheless be read with caution, since the area-specific $p$-values are not adjusted for multiple comparisons and none of the individual rejections of $H_2$ would survive a Bonferroni correction for the eight areas (threshold $0.05/8\approx0.006$). However, three or more such rejections among eight independent tests, performed on disjoint groups of researchers, would occur with a probability of only about $0.6\%$ if $H_2$ held in all areas. Fisher's combined probability test, which merges the $p$-values into a single measure of evidence to which every area contributes, yields a joint $p$-value of about $0.001$, indicating that the three rejections of $H_2$ are unlikely to arise jointly by chance. More fundamentally, the evidence points in the same direction across all areas. $H_1$ is not rejected in any area (even without correction for multiple comparisons), whereas $H_2$ is never favored, with its $\Delta\chi^2$ distribution shifted towards positive values in all eight areas. Consistently, the only inequality pattern pointing in the opposite direction, in Applied Social Sciences, arises where the directional test is inconclusive. Where the evidence is informative, it favors the view that productivity shapes collaboration, with advantage compounding mostly on output rather than on the number of distinct collaborators; where it is not, it is silent rather than contrary.

\section*{Discussion and conclusions}

In summary, our findings provide convergent evidence across multiple analyses that, among distinguished Brazilian researchers holding a nationally prestigious research fellowship, the joint distribution of career totals is considerably more consistent with a generative direction in which productivity shapes collaboration than with the reverse. Across the careers of more than 26,000 researchers spanning a broad range of disciplines, we find that the total number of publications and the total number of unique coauthors are strongly associated, but that this association is markedly asymmetric. Collaboration conditioned on productivity exhibits a smooth mean-variance structure consistent with a geometric model, whereas productivity conditioned on collaboration is substantially less regular. We also find that publication counts closely follow a lognormal distribution, and that the broad distribution of coauthors is described by a geometric-lognormal mixture derived from conditional analysis, without directly fitting the distribution of coauthor counts. Our surrogate tests reinforce this picture by showing that the empirical conditional structure is compatible with the direction in which productivity shapes collaboration, but not with the reverse, thus supporting a parsimonious generative model in which heterogeneity in productivity generates the broad heterogeneity observed in collaboration. A parallel asymmetry appears in how the two quantities are distributed over careers, as the inequality of publications within cohorts rises systematically with career age whereas that of coauthors does not, even though collaboration is the more unequally distributed of the two. This indicates that advantage compounds more on how much a researcher publishes than on how many coauthors they accumulate, providing additional support for the direction of the productivity--collaboration relationship that is not based on the conditional structure. 

The area-resolved analyses temper this aggregate conclusion. The direction in which productivity shapes collaboration is never rejected in any area, whereas the reverse direction is systematically disfavored but is rejected only in Exact and Earth Sciences, Engineering, and Biological Sciences, three rejections that do not individually survive a correction for multiple comparisons but are jointly unlikely to arise by chance. These are broadly the areas organized around traditional laboratory and research-group models, where much of the coauthorship arises within an investigator's own group and a researcher's number of coauthors may therefore reflect their own output more directly. In these same areas, productivity inequality grows with career age markedly more clearly than collaboration inequality, as would be expected if network size largely follows from output. In the remaining areas, neither direction can be sharply distinguished, and the inequality patterns weaken or reverse accordingly. Collaboration inequality grows at least as clearly as productivity inequality in Agricultural and Applied Social Sciences, where more collective authorship practices plausibly allow collaboration to accumulate partly independently of individual output, while neither inequality grows clearly in Linguistics, Literature and the Arts or in Humanities and Human Sciences, where collaboration is often more limited and irregular. Health Sciences is an exception, combining the inequality signature of the laboratory-based areas with an inconclusive directional test, a pattern that might reflect the multi-center and consortium authorship that coexists there with research-group organization, although we have not tested this possibility directly.

By assessing which generative direction is more consistent with the joint distribution of career totals, our research contributes to a still incipient literature~\cite{lee2005impact, he2009research, ductor2015does, abramo2017relationship, zinilli2026collaborative} suggesting that broad variation in collaboration may arise downstream of differences in researchers' ability to sustain productivity. This result is consistent with cumulative-advantage mechanisms in scientific careers, whereby productivity, visibility, and reputation reinforce one another and help attract additional opportunities for collaboration~\cite{merton1968matthew, petersen2014inequality, agrawal2014collaboration}, further echoing earlier views of science as organized around highly productive cores of researchers embedded in broader collaborative structures~\cite{de1966collaboration}. Several channels plausibly sustain this loop: sustained output secures grants, which fund the students and postdocs who themselves become coauthors; it attracts graduate applicants and invitations to join consortia and networks; and it raises the probability of being sought out as a partner, since prospective collaborators may use prior output as a signal of the likely returns to collaborating. We emphasize, however, that this remains a narrative account rather than a formal model. A structural model of collaboration choice, in the spirit of the co-author model of Jackson and Wolinsky~\cite{jackson1996strategic}, in which researchers allocate limited time across collaborations and links form only when both parties expect to benefit, would allow sharper tests than the distributional evidence assembled here and would help to discriminate among the channels listed above, which our data cannot separate, and we regard its development as a valuable direction for future work.

From a research evaluation and science policy perspective, these results indicate a more cautious interpretation of the apparent benefits of collaboration. Because part of the observed collaboration structure appears to be downstream of productivity, the well-documented correlation between collaboration and output should not be read automatically as evidence that expanding collaboration will, by itself, raise output. This is a claim about how existing evidence should be interpreted, and not a recommendation to redirect resources from one policy instrument to another: our results do not imply that collaboration is unimportant, still less that collaborative programs are ineffective, given the well-established benefits of collaboration~\cite{adams2005scientific, freeman2014why}. The disciplinary variation documented in Figures~\ref{fig:5} and~S6 reinforces this caution, since implications drawn from the aggregate would not transfer uniformly across areas. Moreover, the value of collaboration goes beyond the volume of output. Team-authored work has come to dominate knowledge production and attracts more citations~\cite{wuchty2007increasing}, and team size also shapes the character of the research itself, with larger teams tending to develop existing lines of work and smaller ones to disrupt them with new ideas~\cite{wu2019large}. These are dimensions our measures do not capture. Firm policy prescriptions would in any case require the longitudinal and quasi-experimental evidence discussed above, which our design does not provide. More broadly, the findings reinforce the idea that collaboration and scientific productivity should not be interpreted independently but as co-evolving in an asymmetric manner, since the collaboration structure may be partly generated by the unequal distribution of productivity across careers.

Our design should also be situated relative to the approaches that would be needed to identify a causal effect in the potential-outcomes sense. By reducing each career to two aggregate quantities, our directional analysis compresses the temporal information contained in the publication record, and it therefore cannot establish temporal precedence, exploit exogenous variation, or estimate an intervention-level effect. Such totals capture broad regularities, but they cannot accommodate the possibility that different researchers follow distinct regimes, that the direction of the relationship changes across career stages, or that some scientists exhibit stronger bidirectional feedback than others, so that the population-level asymmetry identified here may coexist with substantial individual-level heterogeneity that our framework cannot resolve. Designs based on quasi-natural experiments are considerably more decisive in these respects, as illustrated by the study of the sudden loss of superstar collaborators and its consequences for the subsequent output of their coauthors~\cite{azoulay2010superstar}, and by work reconceptualizing scientific stars in terms of their helpfulness to collaborators rather than their own productivity~\cite{oettl2012reconceptualizing}. Other informative alternatives include matching methods~\cite{rosenbaum1983central} and Granger causality applied to panel data~\cite{granger1969investigating}, as well as nonlinear state-space methods such as convergent cross mapping, which is explicitly designed to infer causal direction from coupled time series~\cite{sugihara2012detecting}. Applying such methods to per-researcher series of annual output and annual new coauthors is, in our view, the natural continuation of this work. These approaches, however, carry their own difficulties when applied to scientific careers. Time-series methods typically require on the order of a hundred annual observations, considerably longer than individual careers can provide: per-researcher trajectories span at most a few decades in our sample, with a median career age of 24 years since PhD and a non-negligible fraction of sparse trajectories. Aggregating across researchers to treat individual series as replicas of a common underlying process is appealing, but it also discards individual-level variation and rests on the strong assumption that careers across researchers, and possibly across fields, are governed by the same dynamics and are well described as a dynamical system. The present study therefore adopts a population-level approach, asking which generative direction better accounts for the joint distribution of career totals across an entire national cohort, and we regard it as complementary to longitudinal designs that may follow it.

A further set of considerations concerns how the two quantities are measured. Because the number of distinct coauthors is enumerated from the publication record, the two totals cannot vary independently and must grow together. This constructional link, however, fixes only that joint growth and leaves largely undetermined the number of distinct coauthors associated with a given level of output, which is precisely the conditional structure our analysis interrogates. Nor does it account for the superlinear scaling we observe, since if each publication contributed a fixed complement of new coauthors, the number of distinct coauthors would grow in proportion to output, and because coauthors recur across a researcher's papers, mere enumeration should, if anything, yield sublinear growth. We nonetheless stress that, although our test evaluates both directions, aggregate career totals cannot establish whether collaborating leads to higher subsequent output within an individual career, which is arguably the most policy-relevant form of the question. Our measure of collaboration is also coarse. It quantifies the breadth of a researcher's coauthorship network but not the substance of the relationships that compose it, and it cannot distinguish a coauthor who shaped a study from one who contributed a specific technical service. The propensity to list contributors also varies across fields, institutions, and individual practices. Our area-resolved analyses partially address the disciplinary component of this heterogeneity, and a role-resolved treatment is a natural direction for future work. Productivity, in turn, is quantified by raw publication counts, without weighting for citations or journal quality. Raw counts and citation-weighted measures capture distinct dimensions of scientific careers, activity in one case and influence in the other, and while Allison and Stewart~\cite{allison1974productivity} found the variation of inequality with career age to be comparable for both, whether the same holds for our directional test is a separate question. The finding that the advantage of team-authored work is more pronounced in citations than in volume~\cite{wuchty2007increasing} indeed raises the possibility that unweighted counts understate the contribution of collaboration to high-impact research, so that we cannot exclude that the productivity--collaboration relationship differs for the most influential work.

The composition of our sample also limits what can be concluded from it. Our cohort is restricted to a relatively homogeneous group of established scientists, which reduces the noise associated with sparse or irregular careers and helps reveal regularities that might otherwise remain obscured, but which equally limits the scope of our conclusions, as the same directional pattern may not hold with equal strength among less prominent scientists or in research systems shaped by different institutional structures and evaluation regimes. Selection nonetheless appears more likely to obscure this asymmetry than to generate it. Conditioning on a high composite of positively correlated attributes tends to induce negative dependence among them within the selected sample, since a researcher admitted with more modest output will typically compensate in impact, supervision, or international reach, which runs counter to the positive association we observe. The inequality analysis points the same way, as the growth rate of the publication Gini in our cohort is considerably shallower than the values reported by Allison and Stewart~\cite{allison1974productivity} for samples not selected on scientific performance, consistent with our selection compressing the range of output and attenuating the cumulative-advantage signal rather than manufacturing it. These are arguments about plausibility rather than demonstrations, and whether the same directional structure holds among less selected researchers can be settled only by direct comparison. A related constraint is that factors we do not observe, including institutional prestige, access to funding, career age, mobility, and differences in individual talent or opportunity, may also affect the relation between productivity and collaboration.

These limitations point to promising directions for future work. Replicating the present analysis on broader and less selected populations of researchers would establish directly whether the asymmetry we report is a property of scientific careers more generally or of the established cohort in which we observe it, and incorporating the covariates listed above would indicate whether the relationship remains stable across contexts or instead reflects a mixture of distinct mechanisms. Extending the analysis to impact-weighted measures of output would, in turn, test whether the same direction holds for the most influential research. Despite these limitations, the consistency of the observed asymmetries across conditional statistics, marginal distributions, surrogate tests, and inequality patterns, and across areas of knowledge, indicates that, in this cohort of Brazilian fellowship holders, the data are considerably more consistent with productivity shaping collaboration than with the reverse.

\subsection*{Competing interest}
The authors declare that they have no competing interests.\vspace{-0.3cm}

\subsection*{Funding}
We acknowledge the support of the Coordenação de Aperfeiçoamento de Pessoal de Nível Superior, the Conselho Nacional de Desenvolvimento Científico e Tecnológico (CNPq - Grants 303533/2021-8, 317522/2025-6, 303765/2017-8, and 302375/2026-0), the Fundação de Amparo à Pesquisa do Estado de São Paulo (FAPESP - Grants 2026/00886-9 and 2023/14427-8), the Fundação Cearense de Apoio ao Desenvolvimento Científico e Tecnológico (FUNCAP), and the Slovenian Research and Innovation Agency (Javna agencija za znanstvenoraziskovalno in inovacijsko dejavnost Republike Slovenije) (Grant P1-0403). The funders had no role in study design, data collection and analysis, decision to publish, or preparation of the manuscript.\vspace{-0.3cm}

\subsection*{Author contributions}
D.A.F., C.I.N.S., V.H.R., H.A.C., M.P., H.V.R., and J.S.A. contributed to the conceptualization, data curation, methodology, software, investigation, validation, formal analysis, visualization, writing–original draft, and writing–review \& editing. G.F.C.L. contributed to the data curation, software, and writing–review \& editing.\vspace{-0.3cm}

\subsection*{Data availability}
The data that support the findings of this study are available as a supplementary file (File S1).\vspace{-0.3cm}

\bibliography{references.bib}



\section*{Supplementary material}

# Supplementary material for "Causal asymmetry suggests productivity underlies scientific collaboration"

**Diego A. Frota,[1,2] Cesar I. N. Sampaio Filho,[1,3] Vítor H. Ribeiro,[4] Germano F. C. Luz,[1,3] Humberto A. Carmona,[1,3] Matjaž Perc,[5,6,7,8,*] Haroldo V. Ribeiro[4,9,*] and José S. Andrade Jr.[1,3,*]**

[1]Departamento de Física, Universidade Federal do Ceará, 60451-970, Fortaleza, Ceará, Brazil, [2]Instituto Federal de Educação, Ciência e Tecnologia do Ceará, Sobral, Ceará, Brazil, [3]Escola de Saúde Pública do Ceará, 60165-090, Fortaleza, Ceará, Brazil, [4]Departamento de Física, Universidade Estadual de Maringá, Maringá, PR, 87020-900, Brazil, [5]Faculty of Natural Sciences and Mathematics, University of Maribor, Koroška cesta 160, 2000, Maribor, Slovenia, [6]Community Healthcare Center Dr. Adolf Drolc Maribor, Ulica talcev 9, 2000, Maribor, Slovenia, [7]Department of Physics, Kyung Hee University, Dongdaemun-gu, Seoul, 02447, Republic of Korea, [8]University College, Korea University, Seongbuk-gu, Seoul, 02841, Republic of Korea and [9]Instituto de Ciências Matemáticas e de Computação, Universidade de São Paulo, São Carlos, SP, Brazil

*To whom correspondence should be addressed. Email: soares@fisica.ufc.br or hvribeiro@uem.br or matjaz.perc@gmail.com



## Abstract

This file contains supplementary figures (S1-S4), tables (S1, S2, and S3) and a file (S1) supporting the manuscript.



## Supplemental Materials

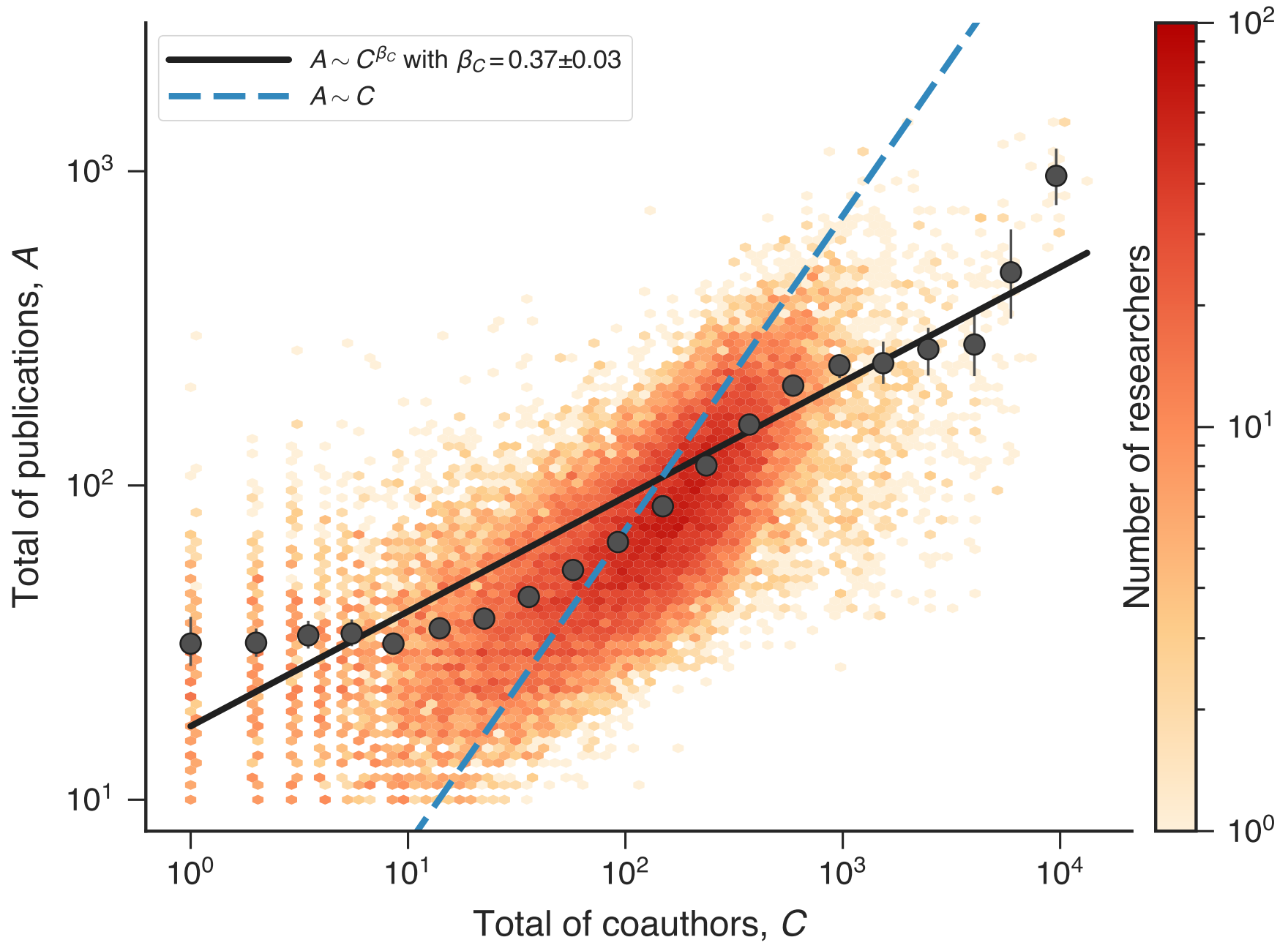


Fig. S1: Relationship between scientific productivity and collaboration under the reverse conditioning. Hexagonal binning map of the joint distribution of the total number of unique coauthors ($C$) and the total number of publications ($A$) for all researchers in our dataset. Both axes are shown on logarithmic scales, and hexagons are color-coded by the number of researchers in each bin. Grey circles denote the average value of $A$ within logarithmically spaced bins of $C$, with error bars representing 99% bootstrap confidence intervals. In contrast to the smooth superlinear trend observed for the average number of coauthors conditioned on publication output (Figure 1 in the main text), the average number of publications conditioned on the number of coauthors exhibits a less regular and weakly structured dependence on $C$, indicating an asymmetric relationship between productivity and collaboration.

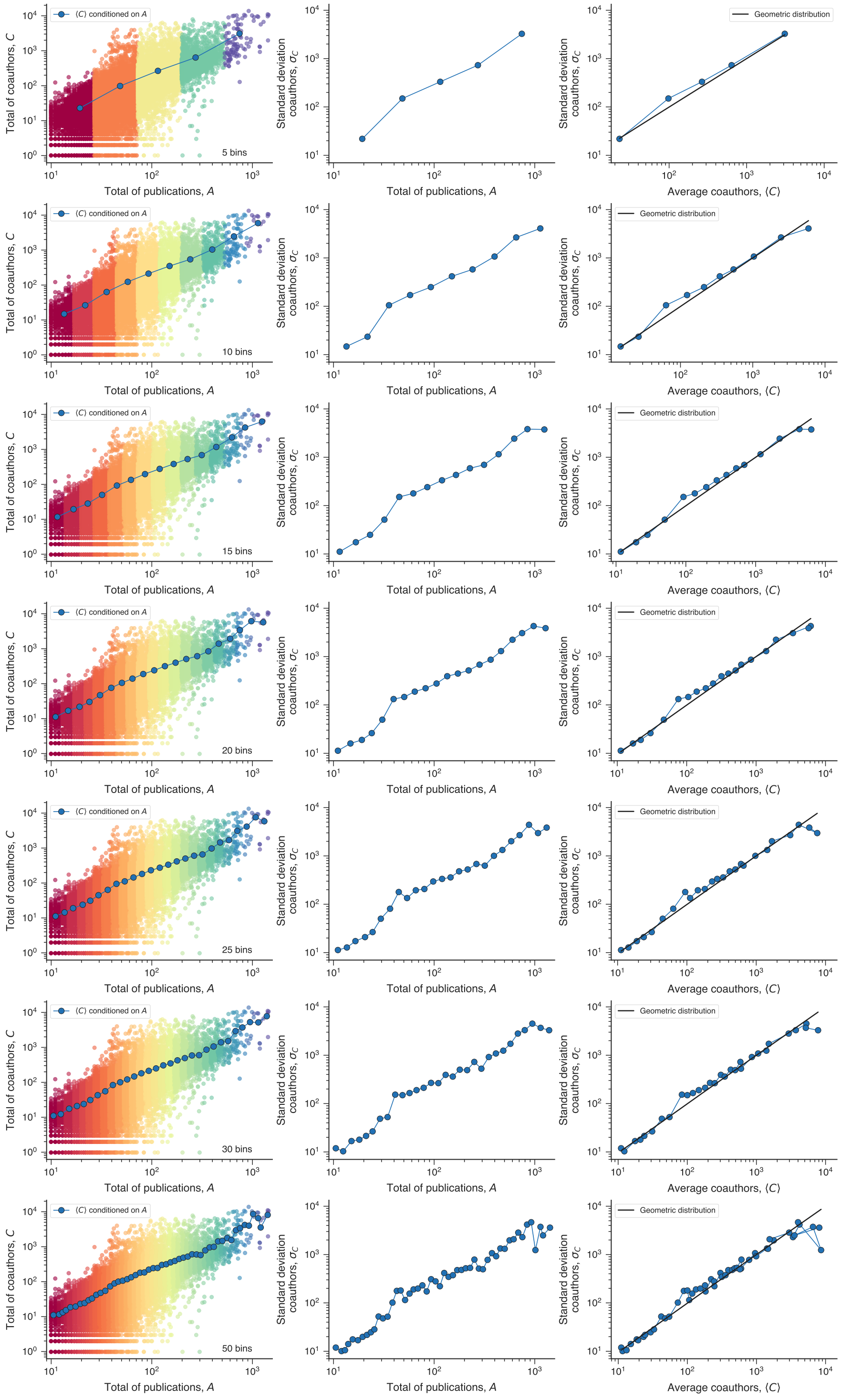


Fig. S2: Robustness of the conditional statistics of collaboration given productivity to the number of bins. Conditional statistics obtained by conditioning on the total number of publications $A$ using 5, 10, 15, 20, 25, 30, and 50 logarithmically spaced bins, arranged from top to bottom by row. In each row, the first column shows the average number of coauthors $\langle C\rangle$ as a function of $A$; the second column shows the standard deviation of the number of coauthors $\sigma_C$ as a function of $A$; and the third column shows the relation between $\sigma_C$ and $\langle C\rangle$, with the solid black line indicating the mean-standard deviation relation expected for a geometric distribution. Across all binning choices, the same qualitative pattern is observed: both $\langle C\rangle$ and $\sigma_C$ increase smoothly with publication output $A$, and the relation between standard deviation and mean remains narrow and close to the geometric expectation.

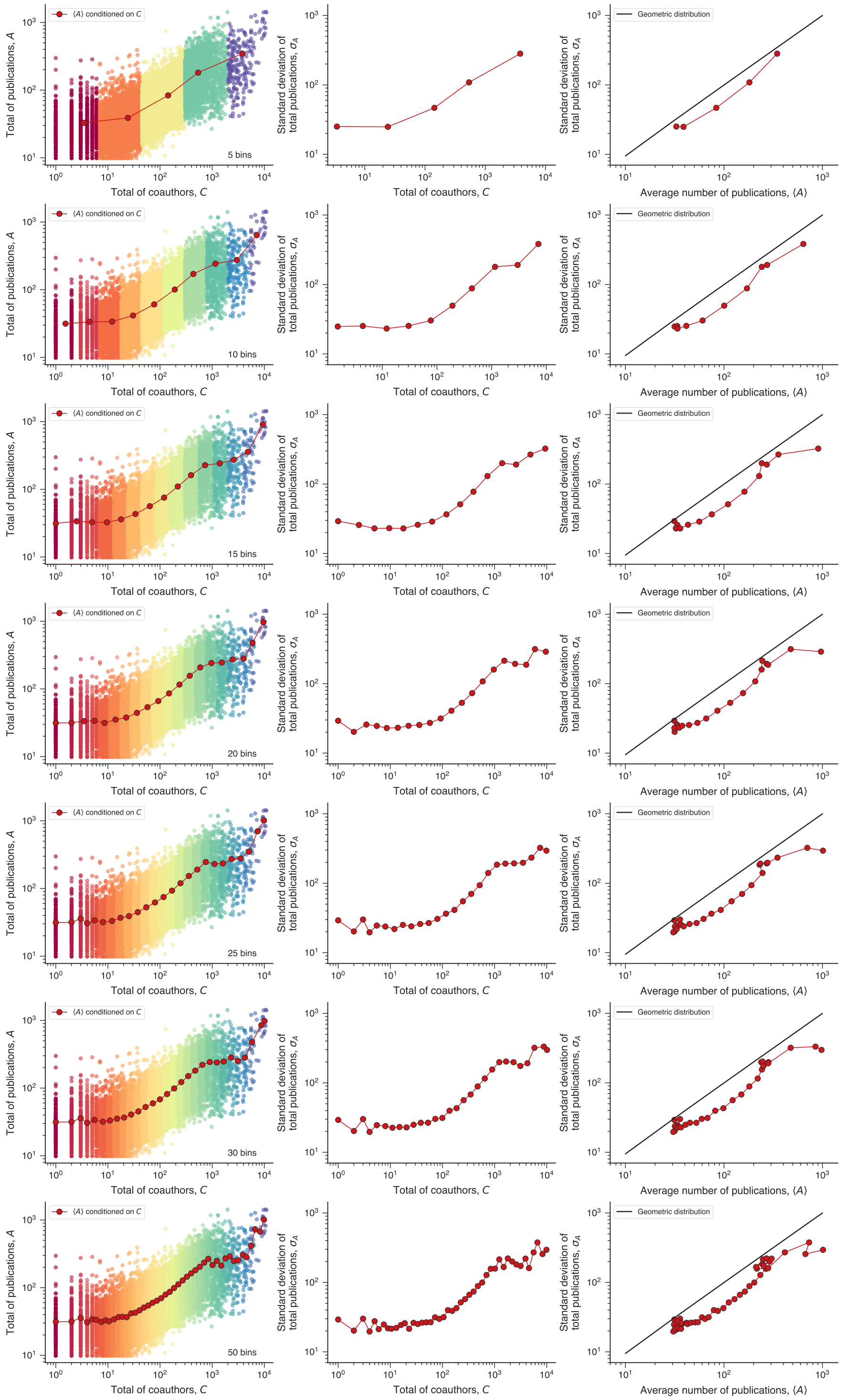


Fig. S3: Robustness of the conditional statistics of productivity given collaboration to the number of bins. Conditional statistics obtained by conditioning on the total number of coauthors $C$ using 5, 10, 15, 20, 25, 30, and 50 logarithmically spaced bins, arranged from top to bottom by row. In each row, the first column shows the average number of publications $\langle A \rangle$ as a function of $C$; the second column shows the standard deviation of the number of publications $\sigma_A$ as a function of $C$; and the third column shows the relation between $\sigma_A$ and $\langle A \rangle$, with the solid black line indicating the mean-standard deviation relation expected for a geometric distribution Across all binning choices, the same qualitative pattern is observed: the conditional statistics of productivity given collaboration remain less regular than those obtained under conditioning on publications, with relatively flat behavior at low $C$ and systematic deviations from the geometric expectation in the relation between standard deviation and mean.

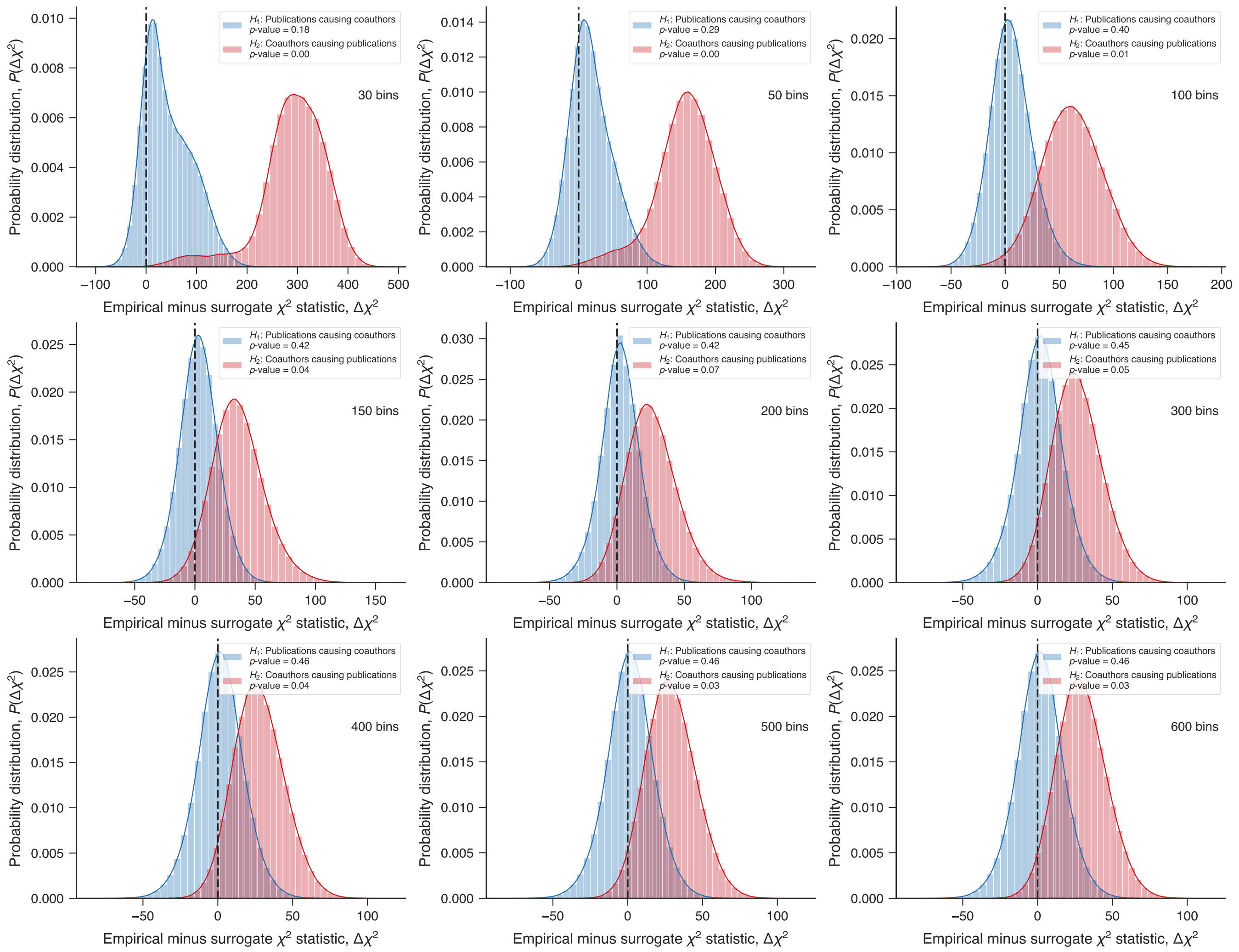


Fig. S4: Robustness of the surrogate-based directional test to the number of bins. Distributions of the empirical-surrogate $\chi^2$ difference, $\Delta\chi^2 = \chi^2_{\mathrm{emp}} - \chi^2_{\mathrm{surr}}$, obtained from the surrogate test for the two competing directional hypotheses using 30, 50, 100, 150, 200, 300, 400, 500, and 600 percentile-based bins, arranged from left to right and top to bottom. In each panel, the blue distribution corresponds to $H_1$, in which productivity drives collaboration, whereas the red distribution corresponds to $H_2$, in which collaboration drives productivity; the corresponding number of bins as well as the one-sided $p$-values are indicated within each panel. Across all binning choices, the distribution under $H_1$ remains centered near zero, indicating that the empirical conditional structure of collaboration given productivity is well reproduced by the surrogate model, whereas the distribution under $H_2$ remains systematically shifted towards positive values, indicating that the empirical conditional structure of productivity given collaboration is not adequately reproduced by the corresponding surrogate model. These results show that the conclusion of the surrogate test is robust against the number of bins.

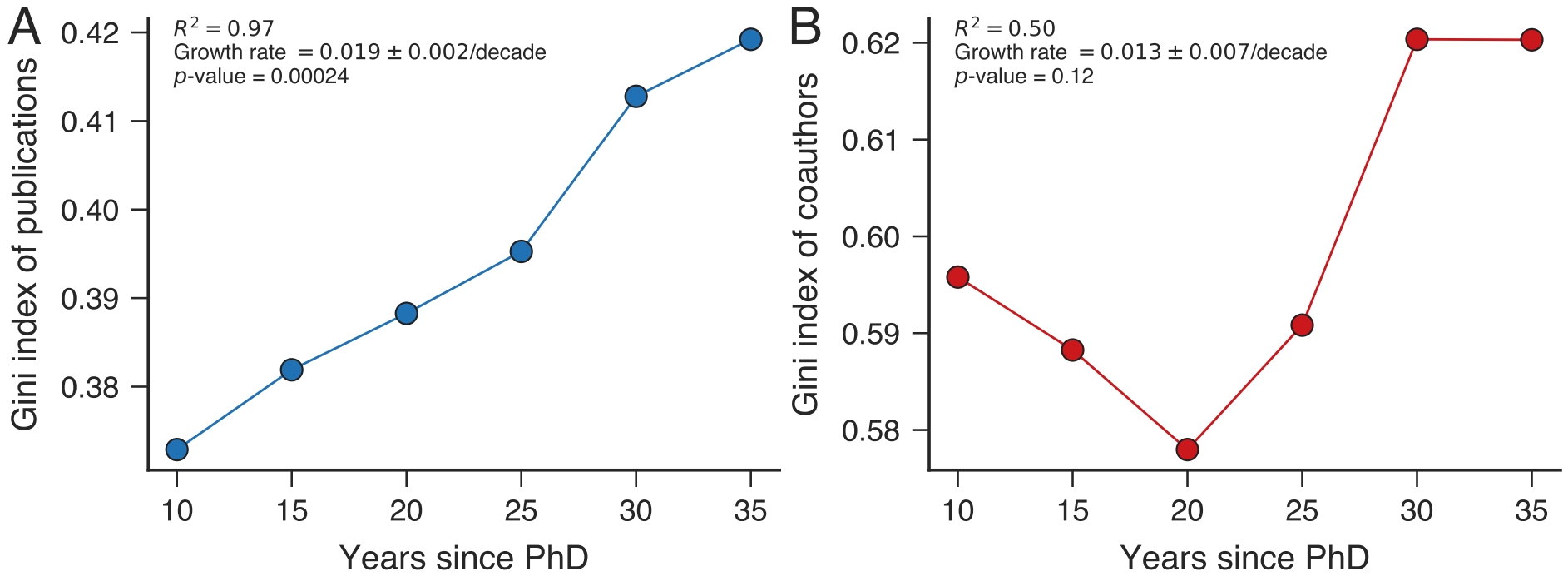


Fig. S5: Inequality of productivity and collaboration as a function of career age. Gini index of (A) the total number of publications, $A$ , shown as blue circles, and (B) the total number of distinct coauthors, $C$, shown as red circles, both evaluated within groups of researchers of similar career age, measured as the number of years since PhD and binned in intervals of five years. Each group is treated as an approximation of a cohort observed at a given career stage, so that the variation across groups indicates whether inequality accumulates over the course of a career. The coefficient of determination $R^2$, the rate of increase per decade, and the corresponding $p$-value of the linear association are indicated in each panel. The inequality of publications increases monotonically and approximately linearly with career age, whereas the inequality of collaboration is substantially higher in level, so that coauthors are the more unequally distributed of the two quantities, but varies far less regularly and non-monotonically, decreasing between 10 and 20 years, increasing between 20 and 30 years, and saturating thereafter. The signature of accumulative advantage is thus considerably stronger for productivity than for collaboration.

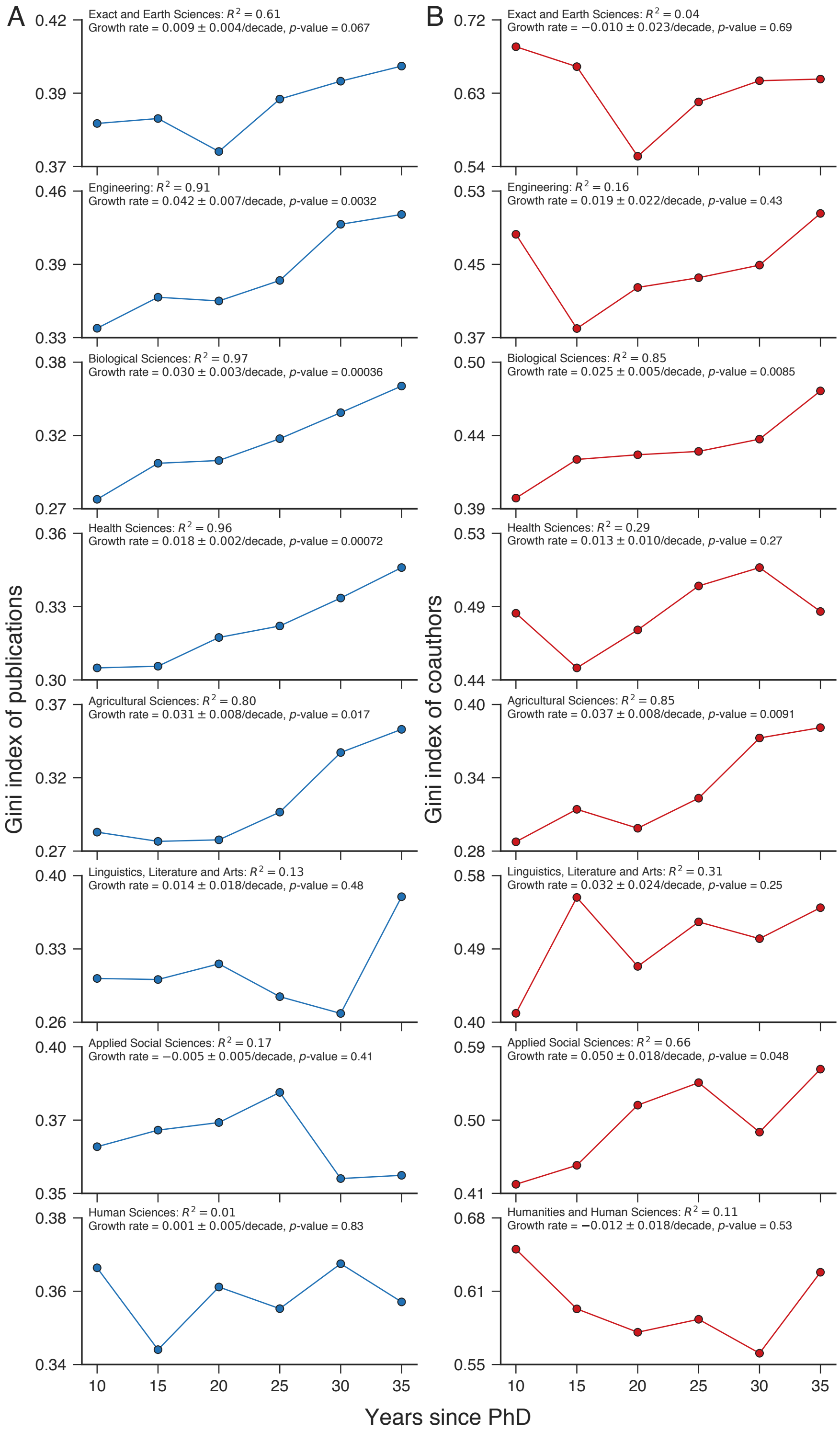


Fig. S6: Inequality of productivity and collaboration as a function of career age, resolved by broad area of knowledge. Gini index of (A) the total number of publications, $A$, shown as blue circles, and (B) the total number of distinct coauthors, $C$, shown as red circles, both evaluated within groups of researchers of similar career age, measured as the number of years since PhD and binned in intervals of five years, and computed separately for each broad area of knowledge. Each row corresponds to one area, indicated within the panels together with the coefficient of determination $R^2$, the rate of increase per decade, and the corresponding $p$-value of the linear association. The inequality of publications increases with career age markedly more clearly than that of coauthors in Exact and Earth Sciences, Engineering, Biological Sciences, and Health Sciences. The two increase comparably in Agricultural Sciences, the ordering is inverted in Applied Social Sciences, where the inequality of coauthors is the one that increases more clearly, and neither shows a clear association with career age in Linguistics, Literature and the Arts or in Humanities and Human Sciences.

**Table S1.** Comparison of candidate distributions for the marginal distribution of productivity. Maximum-likelihood fits of the candidate distributions considered for the total number of publications $A$. The table reports the Kolmogorov-Smirnov statistic ($D_{\mathrm{KS}}$) and the likelihood-ratio quantity relative to the lognormal model. Among all candidates, the lognormal yields the smallest KS statistic and provides the best overall description of the distribution of productivity.

| Distribution | $D_{\mathrm{KS}}$ | Likelihood-ratio |
|---|---|---|
| Lognormal | 0.0085 | 0.00 |
| Lower-truncated power law | 0.0181 | 2.30 |
| Generalized gamma | 0.0213 | 218.70 |
| Inverse Gaussian | 0.0233 | 72.79 |
| Stretched exponential | 0.0561 | 788.81 |
| Gamma | 0.0595 | 1398.47 |
| Weibull | 0.0654 | 2456.93 |
| Generalized Pareto | 0.1473 | 4130.79 |
| Exponential | 0.1534 | 4180.74 |
| Upper-truncated power law | 0.2474 | 5892.40 |
| Pareto | 0.3019 | 14041.10 |

**Table S2.** Comparison of candidate distributions for the marginal distribution of the number of coauthors. Maximum-likelihood fits of the candidate distributions considered for the total number of unique coauthors $C$. The table reports the Kolmogorov-Smirnov (KS) statistic ($D_{\mathrm{KS}}$) and the associated $p$-value from the goodness-of-fit test. None of the candidate parametric forms passed the KS test. Among the distributions fitted over the full support of the data, the generalized Pareto distribution yielded the smallest $D_{\mathrm{KS}}$. The lower-truncated power law [$p(c) \propto C^{-\alpha_C}$, for $C > C_{\min}$] provided the overall smallest $D_{\mathrm{KS}}$ with an exponent $\alpha_C = 2.72$ but restricted to values above $C_{\min} = 257$ coauthors.

| Distribution | $D_{\mathrm{KS}}$ | KS $p$-value |
|---|---|---|
| Lower-truncated power law | 0.0158 | 0.00 |
| Generalized Pareto | 0.0263 | 0.00 |
| Generalized gamma | 0.0332 | 0.00 |
| Stretched exponential | 0.0455 | 0.00 |
| Weibull | 0.0493 | 0.00 |
| Gamma | 0.0647 | 0.00 |
| Lognormal | 0.0658 | 0.00 |
| Exponential | 0.0736 | 0.00 |
| Upper-truncated power law | 0.1754 | 0.00 |
| Inverse Gaussian | 0.2171 | 0.00 |
| Pareto | 0.3560 | 0.00 |
| Lower-truncated generalized Pareto | 0.3700 | 0.00 |

**Table S3.** Median number of publications by broad area of knowledge. For each area, the table reports the number of researchers in our sample and the median number of published articles, $A$. The areas differ considerably in median productivity, but this variation does not follow the strength of the evidence against $H_2$.

| Area of knowledge | Researchers | Median $A$ |
|---|---|---|
| Exact and Earth Sciences | 5,907 | 60 |
| Engineering | 3,331 | 54 |
| Biological Sciences | 4,776 | 82 |
| Health Sciences | 3,268 | 126 |
| Agricultural Sciences | 3,223 | 100 |
| Linguistics, Literature and the Arts | 999 | 36 |
| Applied Social Sciences | 1,973 | 45 |
| Humanities and Human Sciences | 3,301 | 43 |
| All areas | 26,876 | 68 |

**File S1.** The `file_s1.csv` is a CSV file containing 26,876 rows, excluding the header, and four columns: `total_publications`, `total_coauthors`, `field`, and `years_since_phd`. Each row corresponds to a researcher. The columns `total_publications` and `total_coauthors` report, respectively, the total number of publications and the total number of distinct coauthors associated with each researcher. The `field` column indicates the broad area of knowledge to which the researcher is assigned (using integer codes running from 1 to 9 in alphabetical order of the field names), and `years_since_phd` reports their career age group, defined by the number of years elapsed between the PhD and the end of the period analyzed.



\end{document}